\documentclass{WileyMSP-template}
\usepackage{float}

\usepackage{amsmath,amssymb}
\usepackage{booktabs}
\usepackage{array}
\usepackage{ragged2e}
\usepackage{cite}
\usepackage{array} 
\usepackage{booktabs} 
\usepackage{longtable} 
\usepackage{float}
\usepackage{chemformula}
\usepackage{caption} 

\usepackage{soul}
\usepackage{xcolor}
\sethlcolor{white}

\soulregister\cite7
\soulregister\ref7
\soulregister\eqref7
\begin{document}

\pagestyle{fancy}
\justifying

\title{Inverse Design in Nanophotonics via Representation Learning}

\maketitle

\author{Reza Marzban(1)}
\author{Ali Adibi(1)*}
\author{Raphaël Pestourie(2)*}

\dedication{}

\begin{affiliations}(1) School of Electrical and Computer Engineering, Georgia Institute of Technology, Atlanta, GA, USA \\
(2) School of Computational Science and Engineering, Georgia Institute of Technology, Atlanta, GA, USA \\

*Corresponding Authors:  ali.adibi@ece.gatech.edu, rpestourie3@gatech.edu
\end{affiliations}

\begin{abstract}
\textbf{Inverse design} in nanophotonics, the computational discovery of structures achieving targeted electromagnetic (EM) responses, has become a key tool for recent optical advances. Traditional intuition-driven or iterative optimization methods struggle with the inherently high-dimensional, non-convex design spaces and the substantial computational demands of EM simulations. Recently, machine learning (ML) has emerged to address these bottlenecks effectively. This review frames ML-enhanced inverse design methodologies through the lens of \textbf{representation learning}, classifying them into two categories: \textit{output-side} and \textit{input-side} approaches. Output-side methods use ML to learn a representation in the solution space to create a differentiable solver that accelerates optimization. Conversely, input-side techniques employ ML to learn compact, latent-space representations of feasible device geometries, enabling efficient global exploration through generative models. Each strategy presents unique trade-offs in data requirements, generalization capacity, and novel design discovery potentials. Hybrid frameworks that combine physics-based optimization with data-driven representations help escape poor local optima, improve scalability, and facilitate knowledge transfer. \hl{We emphasize data efficiency, transferable representations, fabrication-aware design, faster solvers, and hybrid multiphysics co-design.}
\end{abstract}



\section{Introduction}
\label{sec:intro}
By harnessing subwavelength control of light, nanophotonics has enabled compact imaging systems and displays~\cite{choi2025roll,Chang2024Metalens,Tseng2021NeuralNano,li2022inverse,hemmatyar2021enhanced}, high-throughput optical and neuromorphic-computing platforms~\cite{Camacho2021Parallel,camacho2021single,nikkhah2024inverse,Poordashtban2023CNNMetalines,zarei2020integrated}, high-sensitivity spectroscopic and biochemical sensors~\cite{Tittl2018Barcoding,Hua2022Snapshot,abdollahramezani2024high,dolia2024very}, and emerging architectures for nonlinear, quantum, and reconfigurable photonics~\cite{yang2023inverse,mann2023inverse,abdollahramezani2022reconfigurable,abdollahramezani2022electrically}. The design of nanophotonic devices is fundamentally an \textit{inverse problem}: one specifies a target electromagnetic (EM) response and must infer a compatible physical structure under fabrication and multi-objective constraints. Such inverse problems are ill-posed~\cite{hadamard1902problemes}; outside a handful of highly idealized geometries where Maxwell’s equations admit closed-form inverses, there is no unique or stable mapping from the desired response to the structure~\cite{khaireh2023newcomer}. In realistic settings, the complex interplay of subwavelength geometries, material dispersion, and boundary conditions prevents a tractable analytical solution, leading designers to rely on numerical optimization, an approach complicated by fabrication tolerances that can distort the realized structures~\cite{Sell2019Thesis}. 

Designers typically employ full-wave EM solvers such as the finite-difference time-domain (FDTD)  method~\cite{kunz1993finite} and the finite element method (FEM)~\cite{dhatt2012finite}. Even with graphics processing unit (GPU) acceleration~\cite{zoric2011solving}, large three-dimensional (3D) simulations remain time-consuming and memory-intensive. Solver efficiency and scalability set the practical scope of the design process, as simulation costs often limit the range and resolution of design exploration. This bottleneck is compounded by a broader challenge: neither the optimal device geometry (e.g., the pattern of refractive index) nor optimal physical parameters (such as layer thickness or lattice periodicity) are known \emph{a priori}: they must therefore be optimized~\cite{qian2025guidance}. Traditional manual tuning and heuristic searches are slow and scale poorly in the large design spaces of modern applications~\cite{khaireh2023newcomer,qian2025guidance}. Exhaustive trial-and-error is infeasible; only a small subset of candidate designs can be evaluated within realistic time and memory budgets~\cite{molesky2018inverse}.
This motivates our use of \emph{representation learning} as a conceptual lens: we classify machine learning (ML)-enhanced inverse-design methods not by the application domain or optimizer, but by \emph{where} ML is applied.  \textbf{Output-side representation learning} trains surrogates or end-to-end networks that approximate the forward or inverse EM mapping and replace a part of the physics solver.  \textbf{Input-side representation learning} instead learns a low-dimensional design prior, a latent manifold capturing the salient features of manufacturable high-performance geometries, thus reshaping the search domain.  Viewing the field through this solver-versus-geometry lens complements existing ML-in-photonics surveys~\cite{khatib2021deep,campbell2023advances,hao2022physics,wiecha2021deep,kudyshev2020machine,jiang2021deep} and highlights opportunities for hybrid and transferable design pipelines.

\begin{figure}[H]
    \centering
    \includegraphics[width=0.95\linewidth]{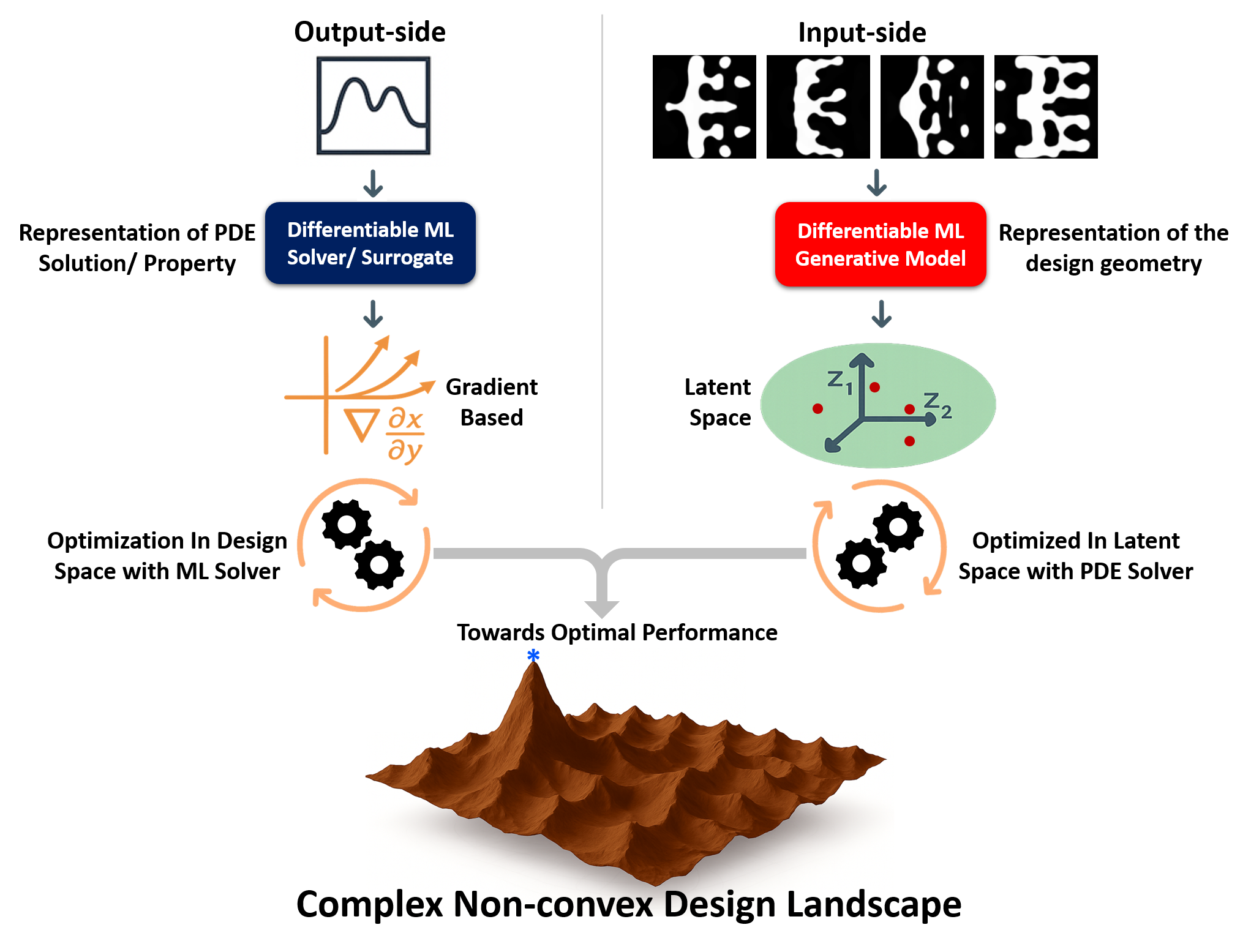}
\caption{\textbf{Output-side versus input-side representation learning in nanophotonic inverse design.}  
Top panels: two complementary learned representations. 
\textbf{Output-side representation} (left) models the \emph{partial-differential-equation (PDE) solution or a derived optical property}: a differentiable surrogate or physics-informed neural network (PINN) emulates a Maxwell’s equations solver and provides analytic gradients that refine candidates directly in the full design space.  
\textbf{Input-side representation} (right) models the \emph{device geometry itself}: a generative model compresses layouts into a low-dimensional latent manifold, and optimization proceeds in that manifold while a PDE solver (the surrogate model) supplies the objective.  
Bottom panel: schematic of the non-convex nanophotonic design landscape.
Two representation-learning frameworks guide the search toward a better optimum (blue star): the \textbf{output-side} surrogate delivers rapid, physics-consistent gradients, and the \textbf{input-side} latent prior confines exploration to geometry regions containing high-performance candidates.
Taken separately, each paradigm yields faster convergence, fewer full-wave EM simulations, and lower data requirements; their distinct mechanisms and trade-offs are analyzed in detail throughout this review.}
\label{fig:evolution}
\end{figure}

To place the representation-learning framework in context, we first review the physics-driven optimization methods that have shaped nanophotonic inverse design. Foremost are gradient-based techniques especially topology optimization (TO)~\cite{qian2025guidance,ji2023recent}—which have become the cornerstone of nanophotonic design~\cite{Bendsoe2013Topology,Christiansen2021Tutorial,Lin2019TopoMS,Sell2019Thesis,pestourie2025fast}. They can handle thousands of design variables and generate complex non-intuitive designs that outperform traditional ones~\cite{molesky2018inverse}. A standard TO loop defines an optical figure of merit (FoM) for the desired optical performance and then iteratively updates the refractive-index distribution to maximize (or minimize) the defined objective~\cite{Bendsoe2013Topology,Christiansen2021Tutorial,Lin2019TopoMS,pestourie2025fast}. Because a single adjoint simulation provides the full gradient, each iteration scales independently of the parameter count~\cite{pestourie2020assume}. Recent differentiable-solver frameworks go further: they compute these gradients directly during each iteration, so no extra adjoint simulations are needed\hl{~\cite{colburn2021inverse,deng2021neural}}. 

As the complexity of the optimization problem grows (e.g., more degrees of freedom, multi-objective optimization), poor local optima may hinder the optimization process~\cite{khaireh2023newcomer}. As a result, the final outcome may depend more strongly on the initial guess, and globally optimal solutions may become elusive. Furthermore, while the gradient calculation itself can be efficient, the iterative nature of these methods, requiring repeated full-wave EM simulations, can lead to prohibitive computational costs and scalability issues, especially for large-scale 3D structures or multi-objective problems. Practical limits, such as structure diameters of approximately $500\lambda$($\lambda$ wavelength), have been observed, with simulations for centimeter-scale devices demanding terabytes of memory and years of computation time~\cite{kang2024large}. Most of the iterative effort of gradient-based optimization is spent on gradient steps: the simulations and gradient evaluations are only used to determine the next guess in the optimization until a local optimum is reached. Standard gradient-based workflows are memoryless: after reaching a local optimum, they do not leverage the record of the search path that could help a later run. All optimizations have to start from scratch, wasting previous efforts and preventing any meaningful knowledge reuse or transfer learning. Density-based TO also tends to yield a final structure with limited insight into the underlying physical mechanisms or more sophisticated knowledge, like the robustness of the design to fabrication imperfections. Many TO workflows assume fixed physical hyperparameters, such as thicknesses of the different material layers or the period of a periodic structure, from the outset, leaving potentially superior designs unexplored. Shape-based parameterizations can vary some physical parameters and produce simpler layouts~\cite{li2022inverse,gershnabel2022reparameterization,dainese2024shape,panuski2022full}, but the reduced degrees of freedom may cap ultimate performance~\cite{pestourie2025fast}.

\hl{Beyond these parameterization choices, the additional degrees of freedom in topology optimization come at the cost of fabrication constraints imposed during the design process. For planar devices, manufacturability is typically ensured through projection or filtering schemes that regulate feature sizes and reduce sensitivity to process variations~\cite{Bendsoe2013Topology}. Recent advances, including foundry-compatible workflows~\cite{hiesener2025seeded}, demonstrate that high-performance, fabrication-ready layouts can now be realized within standard lithographic constraints. In free-form three-dimensional architectures, additional limitations arise from structural integrity, which requires coupling electromagnetic optimization with mechanical stability models~\cite{augenstein2020inverse,kuster2025inverse}. Furthermore, extending topology optimization to nonlinear systems introduces new challenges. Although the adjoint method is inherently linear, such cases mainly require deriving a problem-specific adjoint solver~\cite{hughes2018adjoint}.}

Alongside physics-based optimization, purely data-driven methods have appeared. Neural networks (NNs)~\cite{peurifoy2018nanophotonic,so2020deep,Kiarashinejad2020DimRed,kiarashinejad2020knowledge,kiarashinejad2019deep,zandehshahvar2022manifold,An2019Objective,An2020Deep}, generative models~\cite{Jiang2019GLOnet, Liu2018Generative,Wen2020GAN,kudyshev2020machine}, and reinforcement learning (RL) agents~\cite{li2023deep,park2024sample,sajedian2019optimisation}, have been proposed to either work on the input side (suggesting high-performing photonic layouts) or on the output side (rapidly predicting the optical response of a device). An ML inference is far faster than that by a full-wave EM solver. However, in practice, amortizing the training costs of ML algorithms may defeat their purpose. First, training an accurate model requires a large and high-quality dataset. Naive sampling of the design space at random will flood the training set with mainly poor-performing structures, wasting expensive simulations on unpromising examples ~\cite{peurifoy2018nanophotonic}. Indeed, most deep NNs still require thousands of labeled device examples ~\cite{khaireh2023newcomer} with reasonable response, and generating those labels via full-wave 3D simulations is often computationally prohibitive ~\cite{liu2018training}. Even a well-trained NN may fail on designs or target conditions outside its training distribution~\cite{wiecha2021deep,chen2022high}. Moreover, the inverse mapping from a desired response to a device geometry is generally \emph{one-to-many} (multiple distinct structures can exhibit nearly indistinguishable responses), which complicates direct geometry predictions~\cite{frising2023tackling}. 

These limitations have spurred the development of \emph{hybrid} inverse design strategies~\cite{wang2022advancing}. Rather than relying exclusively on local gradient updates or purely data-driven exploration, hybrid frameworks integrate the key features of the two to address their existing challenges. For example, an ML model can generate diverse initial guesses. An adjoint routine then refines them~\cite{Jiang2019GLOnet}. Likewise, a global optimizer (e.g., a genetic algorithm (GA)~\cite{mitchell1998introduction}) can run alongside a fast surrogate trained on physics-based data~\cite{wu2024localized,zhang2022inverse}. RL~\cite{sutton1998reinforcement} agents explore large design spaces with reward signals derived from Maxwell’s equations~\cite{li2023deep}, and dimensionality-reduction techniques like variational autoencoders (VAEs) can project device geometries into latent spaces more amenable to optimization~\cite{marzban2025hilab}. By combining these tools, hybrid methods escape local minima, reduce solver calls, and reuse prior knowledge on new tasks. As hybrid approaches mature, they fill the gap between local precision and global exploration. Because they merge physics-driven and data-driven strengths, hybrid methods tackle key issues, including scalability, robustness, and knowledge transfer, that each class of design methods struggles to solve alone.

\hl{Overall, the evolution of inverse design in nanophotonics points to a complexity that neither pure physics-based methods nor pure data-driven models can tackle alone. Previous surveys have established a strong foundation by linking ML and photonics. Early work, for example, categorized supervised surrogates and end-to-end inverse pipelines for metamaterials and metasurfaces~\cite{khatib2021deep,campbell2023advances}. Others critically appraise nanophotonic inverse design through the lens of deep NNs and generative models~\cite{wiecha2021deep,jiang2021deep}. Complementary work surveys ML-assisted global optimization schemes~\cite{kudyshev2020machine} and physics-informed ML in EM and inverse problems more broadly~\cite{hao2022physics}. Finally, ref~\cite{woldseth2022use} examined synergies between ML and TO.

In essence, these valuable overviews primarily organize the literature either by application domain~\cite{khatib2021deep,campbell2023advances,molesky2018inverse,qian2025guidance} or by the machine-learning model class employed~\cite{wiecha2021deep,jiang2021deep,kudyshev2020machine,hao2022physics,woldseth2022use}. This review complements these perspectives by classifying methods through the lens of \textbf{representation} \textbf{learning}: ML either targets the \emph{output} space, using surrogates to emulate solver responses, or targets the \emph{input} space, using generative or latent models to parameterize geometry. This solver-versus-geometry approach provides a shared vocabulary for comparing disparate techniques and exposes fundamental trade-offs in data efficiency and generalization. While we present these classes separately for clarity, their power is also realized in hybrid frameworks where input-side generative priors are coupled with output-side surrogate solvers.}

This paper is structured as follows. Section~\ref{sec:classeA} reviews output-side representation learning methods (Class~A), emphasizing differentiable surrogate solvers. Section~\ref{sec:classeB} introduces input-side approaches (Class~B), highlighting latent-space geometric representations. Section~\ref{sec:classeC} describes how these techniques integrate with global optimization algorithms and duality-based performance bounds. Finally, Section~\ref{sec:future} discusses current challenges and suggests future research directions in hybrid inverse-design methodologies.

\section{Class~A: Differentiable Solvers and Surrogate Models (Output-Side Representation Learning)}
\label{sec:classeA}

\emph{Output-side representation learning} forms the first class. This paradigm leverages ML techniques to learn a forward or inverse mapping, typically actualized through surrogate models. The core objective is to emulate, augment, or accelerate the physical solver itself (see the left branch of Figure~\ref{fig:evolution}). In this context, the representation being learned pertains to the response of the physical system (e.g., EM fields or transmission spectra) or the behavior of the solver, rather than a representation of the device geometry. The surrogate’s gradients with respect to all design variables enable the optimizer to navigate the highly non-convex landscape shown at the bottom of Figure \ref{fig:evolution}. These methods build upon classical physics-driven optimization principles but strategically integrate ML to enhance the efficiency, scope, or differentiability of the process by which a device performance (the output or the response) is predicted or its gradients with respect to design parameters are obtained.
The foundation of this class lies in established physics-driven optimization techniques, that leverage gradient-based optimization~\cite{Sell2019Thesis,Bendsoe2013Topology,Christiansen2021Tutorial,Lin2019TopoMS,pestourie2025fast}. The important novelty and the defining feature of this class in recent years is the innovation with ML techniques on the \emph{solver} side, or more broadly, the \emph{output side} of the PDE solution. This involves training ML models to predict EM fields, approximate solutions to Maxwell's equations, or function as rapid, differentiable surrogates for computationally intensive full-wave EM solvers. Several frameworks exemplify the application of output-side representation learning. Recent advancements include the development of end-to-end differentiable EM simulators, where the entire simulation pipeline, or key parts of it, can be differentiated with respect to design parameters. This enables gradient computation through automatic differentiation, streamlining the optimization process ~\cite{colburn2021inverse,minkov2020inverse,mahlau2025flexible}. While earlier reviews covered differentiable surrogate models~\cite {ji2023recent,kim2025inverse}, we focus on recent hybrid surrogates that involve physics-augmented representation learning.
\begin{figure}[H]
    \centering
\includegraphics[width=1\linewidth]{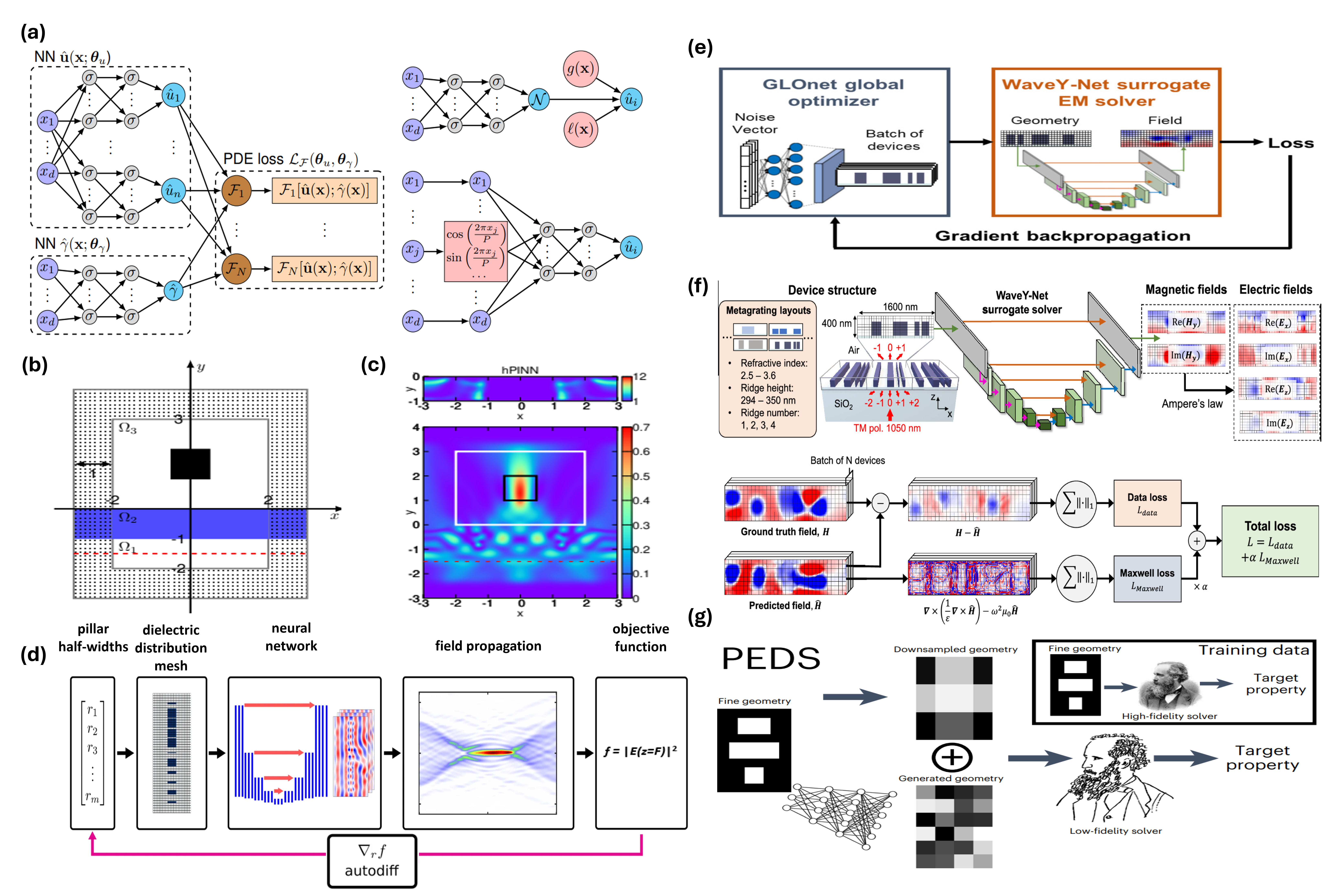}
    \caption{\textbf{Gradient-based co-optimization frameworks combining differentiable EM simulators with PINNs and neural surrogates.}  
    \textbf{(a–c)} \emph{Hard-constrained PINNs:} Two NNs, \(\hat u(\mathbf{x};\theta_{u})\) and \(\hat\gamma(\mathbf{x};\theta_{\gamma})\), parameterize EM fields and design variables. Training involves a PDE-informed loss function \(\mathcal{L}_{F}\) imposed through automatic differentiation. Dirichlet boundary conditions are enforced in the network outputs, and periodic boundary conditions are embedded via sinusoidal input features.  
    (b) Computational domain showing permittivity design region \(\Omega_{2}\) (blue) and perfectly matched layers (PML, hatched).  
    (c) Predicted electric field intensity distribution \(|E|^{2}\) resulting from the optimized permittivity \(\varepsilon\).  
     \textbf{(d)} \emph{Neural-adjoint patch solver:} Pillar half-width vectors are transformed into dielectric patches, processed by a convolutional NN (CNN) predicting local EM fields. These fields are stitched together and propagated via the angular-spectrum method. The objective intensity \(f=|E(z=F)|^{2}\) is back-propagated using automatic differentiation (PyTorch autograd) to iteratively update the pillar geometries.
    \textbf{(e,\,f)} \emph{GLOnet + WaveY-Net framework for global TO:}  
    (e) GLOnet generates metagrating designs from latent noise vectors; these designs are evaluated by the differentiable WaveY-Net surrogate EM solver. Loss gradients computed by WaveY-Net are back-propagated through GLOnet, enabling differentiable, end-to-end global optimization.  
    (f) WaveY-Net architecture details: A U-Net-based CNN predicts magnetic near-fields, subsequently converted to electric fields via the discrete Ampère’s law. The training loss includes a data-fidelity term (\(L_{\text{data}}\)) and a Maxwell-residual regularizer (\(L_{\text{Maxwell}}\)) to keep gradient computations consistent with Maxwell’s equations.  
    \textbf{(g)} \emph{Physics-enhanced deep surrogate (PEDS) framework:} Fine-resolution geometries are downsampled and combined with coarse-resolution geometries generated by a neural surrogate. The resulting composite structures are evaluated by a fast, low-fidelity solver for rapid performance estimation. High-fidelity solver evaluations of fine-resolution geometries provide offline training data, accelerating the design optimization loop while maintaining physical accuracy.
    \medskip
    \textbf{Permissions:}
    Panels (a–c) \textbf{Reproduced with permission.}\textsuperscript{\cite{lu2021physics}}
    \textbf{Copyright 2021, Society for Industrial and Applied Mathematics (SIAM).}
    Panel (d) \textbf{Reproduced under the terms of the CC BY 4.0 license.}\textsuperscript{\cite{zhelyeznyakov2023large}} \textbf{Copyright 2023, The Author(s).}
    Panels (e, f) \textbf{Reproduced with permission.}\textsuperscript{\cite{chen2022high}} \textbf{Copyright 2022, American Chemical Society.}
    Panel (g) \textbf{Reproduced with permission.}\textsuperscript{\cite{pestourie2023physics}} \textbf{Copyright 2023, Springer Nature.}}
    \label{fig:Inloop}
\end{figure}
\noindent
A first approach to physics-augmented learning are PINNs, where governing physical laws, such as Maxwell's equations, are directly embedded into the NN loss function~\cite{raissi2019physics,lim2022maxwellnet}. PINNs can solve the inverse problem in about the same time as the forward problem. Although they are slow forward solvers compared to the state-of-the-art, they can solve the inverse problem competitively~\cite{chen2020physics}.
One example is the PINN with hard constraints (hPINN)~\cite{lu2021physics}, in which the optimal design and PDE solution are simultaneously discovered by solving the inverse design problem directly through a PDE-informed loss. In hPINN, there is no explicit separate parameterization for distinct inputs or design configurations; rather, the network concurrently identifies the optimal geometry and associated EM fields.

This integrated framework inherently constrains the network to solutions consistent with fundamental physics. Figures \ref{fig:Inloop}(a–c) show an hPINN implementation \cite{lu2021physics}. Specifically, two NNs are jointly employed: $u(x;\theta_u)$, representing the EM field, and $\gamma(x;\theta_\gamma)$, representing the material distribution (e.g., permittivity). The physics-informed loss function $\mathcal{L}_F$ penalizes deviations from Maxwell’s equations and explicitly enforces boundary conditions, e.g., including Dirichlet constraints directly imposed on network outputs and periodicity embedded through sinusoidal input features (see Figure~\ref{fig:Inloop}(b)). These jointly trained networks simultaneously yield predictions of the EM field distribution (Figure~\ref{fig:Inloop}(c)) and the corresponding optimized material structure. The hPINN, therefore, learns the solution space defined by Maxwell's equations, realizing output-side representation learning by directly encoding physical fields and material properties.

Another PDE-loss–based physics augmentation integrates PINNs into differentiable surrogate frameworks. For example, Ref.~\cite{zhelyeznyakov2023large} demonstrated a CNN-based PINN surrogate that learns local EM field solutions for metasurface elements (see Figure~\ref{fig:Inloop}(d)). Here, geometric parameters (pillar dimensions) define inputs to the network, which predicts scattered fields. These local field solutions are then stitched via an overlapping-domain approximation method~\cite{lin2019overlapping} and propagated using the angular-spectrum method, yielding a full metasurface response. The CNN–PINN model in figure~\ref{fig:Inloop}(d) effectively learns a differentiable forward solver constrained by Maxwell’s equations, subsequently enabling efficient gradient-based optimization of geometry after the initial training. The important step in this method is backpropagating gradients of the FoM, for instance, the objective intensity ($f = \left| E(z=F) \right|^2$), through the entire differentiable surrogate, including the NN. This enables iterative updates of the device geometry parameters. Thus, the NN effectively learns a differentiable forward solver constrained by Maxwell’s equations, mapping local geometric features directly to corresponding local field responses. This learned representation accelerates full-wave solutions and facilitates efficient, gradient-based optimization of the device geometry following the initial training.

Extending the representation further, pixel-based differentiable surrogates such as WaveY-Net~\cite{chen2022high} use PDE losses to directly learn forward solvers from pixel-level geometric inputs. Unlike pillar-based parameterizations, WaveY-Net encodes metasurface geometries explicitly as pixel patterns. It predicts magnetic near-field distributions, then converts them to electric fields through discrete Maxwell relations. The WaveY-Net training incorporates both a data-fidelity term ($\mathcal{L}_{\text{data}}$), ensuring accuracy with respect to full-wave simulations and a Maxwell-residual regularizer ($\mathcal{L}_{\text{Maxwell}}$) to keep the solution physical. After training, this pixel-based surrogate quickly evaluates device designs, facilitating gradient-based optimization in a fully differentiable manner.

The GLOnet + WaveY-Net framework, illustrated in Figures~\ref{fig:Inloop}(e,f), exemplifies this surrogate approach within global TO of metagratings~\cite{chen2022high}. GLOnet~\cite{Jiang2019GLOnet}, a generative model, proposes candidate designs encoded as pixel patterns, which WaveY-Net then evaluates as a differentiable surrogate EM solver (Figure~\ref{fig:Inloop}(e)). Figure~\ref{fig:Inloop}(f) shows that WaveY-Net adopts a U-Net architecture trained specifically to map pixelized input structures to predicted near-fields. The differentiability of WaveY-Net enables backpropagation of gradients to flow through the entire network and even through GLOnet, allowing for efficient, fully end-to-end optimization. WaveY-Net, therefore, serves as a fast physics-consistent surrogate solver that speeds up evaluation and gradient-based optimization loops.

Beyond explicit PDE-loss approaches, a distinct class of physics augmentation methods involves embedding differentiable approximate solver layers directly within neural frameworks. One example is the physics-enhanced deep surrogate (PEDS) framework~\cite{pestourie2023physics}, shown in Figure~\ref{fig:Inloop}(g). Instead of directly parameterizing device geometry, PEDS learns an optimized \textit{input representation} tailored specifically for a low-fidelity physical solver. In this approach, fine-resolution device geometries are initially downsampled and subsequently combined with coarse-resolution geometries generated by a neural surrogate. These composite structures serve as inputs to the fast approximate solver, improving its accuracy without fundamentally altering the underlying geometric parameterization. The PEDS framework employs offline training data generated through high-fidelity solver evaluations of fine-resolution geometries. Using this data, the neural surrogate learns to approximate key aspects of the system behavior, such as coarse-resolution field properties or critical performance metrics. This learned input representation effectively accelerates the differentiable solver layer, enabling rapid performance estimation. Consequently, the PEDS approach facilitates efficient gradient-based optimization loops by swiftly approximating device performance, thus reducing computational costs compared to conventional full-wave EM simulations.

An important enabling factor for many of these output-side methods is the differentiability of the ML components and, increasingly, the entire simulation and optimization pipeline. The neural-adjoint patch solver relies on automatic differentiation for backpropagation (Figure~\ref{fig:Inloop}(d)~\cite{zhelyeznyakov2023large}). PINNs apply their physics-informed loss via automatic differentiation (Figure~\ref{fig:Inloop}(a-c)~\cite{lu2021physics}). The GLOnet + WaveY-Net framework achieves fully differentiable end-to-end optimization due to the differentiable nature of the WaveY-Net surrogate (Figure~\ref{fig:Inloop}(e,f)~\cite{chen2022high}). The differentiability ensures the flow of gradient information from the final FoM back through the learned output model (and potentially through any generative model proposing the designs) to the design parameters themselves. This capability unlocks gradient-based optimization across complex, ML-augmented systems and signifies a trend towards making increasingly larger portions of the design pipeline differentiable.
Output-side representation learning methods have clear advantages. By providing analytical or automatically differentiated gradients (e.g., TO, differentiable solvers) or through rapid surrogate evaluations, these techniques can navigate high-dimensional design spaces more efficiently than gradient-free searches. The learned representation of the solver's output or its operational behavior is central to this acceleration. They have produced many high-performance photonic devices with complex geometries~\cite{Sell2019Thesis,Bendsoe2013Topology,Christiansen2021Tutorial,Lin2019TopoMS,pestourie2025fast}. Furthermore, the direct incorporation of physical laws in frameworks like PINNs (Figures~\ref{fig:Inloop}(a-c)~\cite{lu2021physics}) or through physics-based regularization in surrogates like WaveY-Net (Figure~\ref{fig:Inloop}(f))~\cite{chen2022high} leads to more robust and generalizable learned models of the system's output. For surrogate models to be truly effective, they must balance computational speed with predictive accuracy. Physics-based regularization helps maintain that fidelity; for instance, WaveY-Net's Maxwell-residual term~\cite{chen2022high} ensures its predictions are physically plausible. This helps to mitigate the black-box concerns often associated with purely data-driven ML models.

Despite their successes, direct gradient-based methods and their ML-augmented counterparts face important limitations. They often stall in local optima because they rely on local gradients in a rugged, non-convex design landscape encountered in nanophotonics. Random restarts help, but they add extra computational cost~\cite{Sell2019Thesis}. Many TO workflows fix parameters such as layer thickness or lattice period. This simplifies the search but can skip unconventional, higher-performance designs~\cite{marzban2025hilab}. \hl{While surrogate models can dramatically accelerate computations, the curse of dimensionality still scales exponentially with the number of design parameters. This acceleration can extend the range of feasible explorations, but asymptotically the exponential growth in data need from the curse of dimensionality dominates, limiting the practical benefit unless the search space is reduced or structured. Moreover, surrogate models are inherently biased toward the distribution of their training data, making knowledge transfer across different material systems, boundary conditions, or target responses difficult. Learning directly from the underlying geometry tends to yield representations that are more transferable than those learned purely from the physics-based outputs, motivating the hybrid and input-side strategies discussed later in this work.}

\section{Class B: Representation-Learning for the design space (Input-Side\\Representation Learning)}
\label{sec:classeB}

The second class of inverse-design methods shifts the focus of ML from the solver or its output to the input or the geometry of the device. We interchangeably use the words input side and design space to denote the feasible space of the PDE-constrained optimization problem. Input-side representation learning seeks a latent manifold that captures the key low-dimensional features of feasible, high-performance structures. Rather than emulating the solver, these methods build a learnable representation of geometry. The objective is to discover low-dimensional manifolds (latent design spaces) that capture the important features of high-performance photonic structures~\cite{melati2019mapping,Zandehshahvar2021DimRed,wang2019clonal,Ma2019Probabilistic}.
\begin{figure}[H]
    \centering
    \includegraphics[width=1\linewidth]{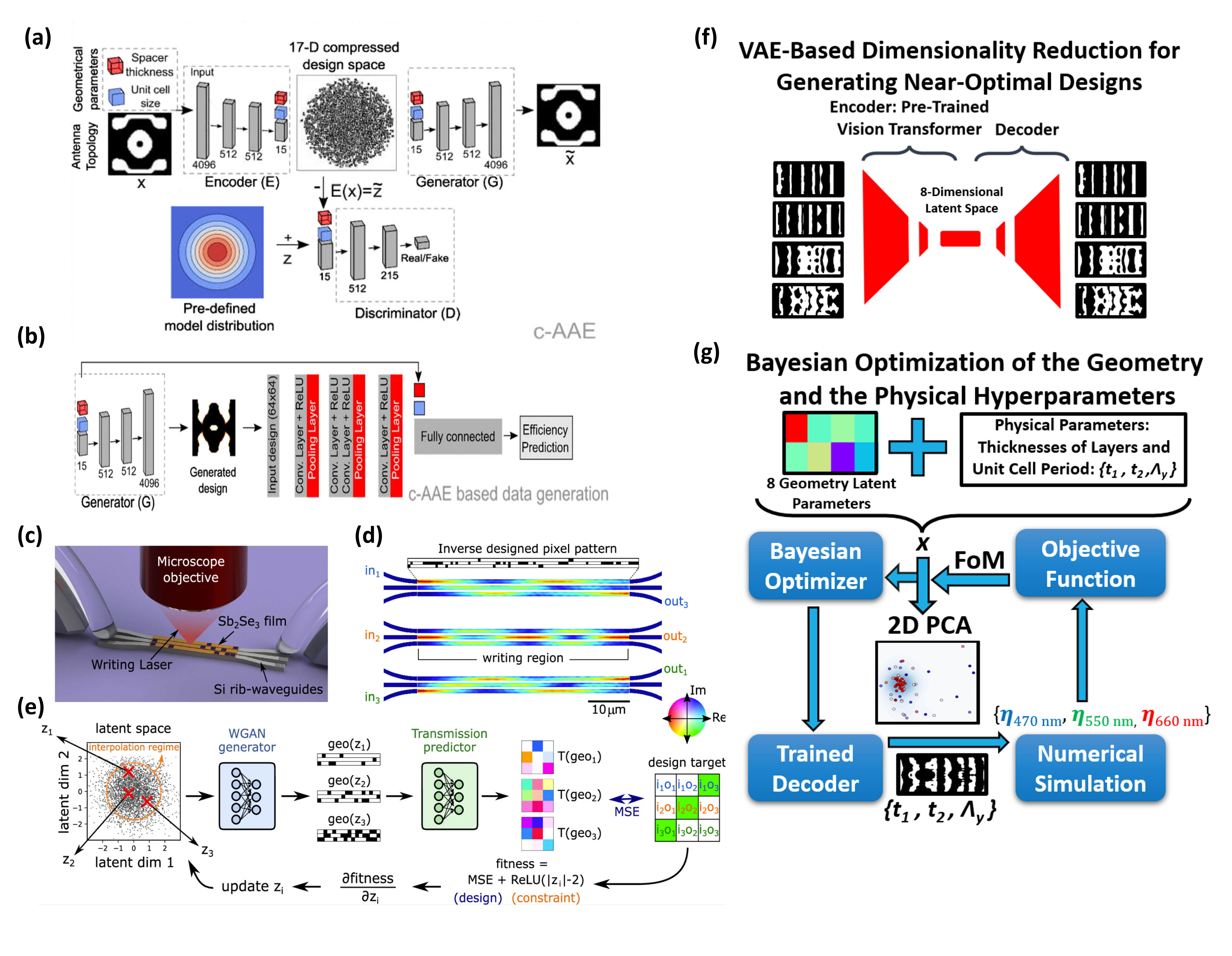}
    \caption{\textbf{Offline, data-driven hybrid inverse-design workflows.}  
    \textbf{(a,\,b) Conditional adversarial auto-encoder (c-AAE) pipeline.}  
    (a) Antenna topology, along with geometric parameters (unit-cell size, spacer thickness), is encoded into a 17-dimensional latent space. A generator–discriminator pair enforces adherence to a predefined prior, producing a compact, physics-informed design manifold.  
    (b) The trained generator \(G\) couples to a conditional VGG-based surrogate model for rapid offline prediction of optical efficiency, allowing rapid synthesis and screening of candidates.  
    \textbf{(c–e) CNN-assisted WGAN optimization for reconfigurable photonic waveguides.}  
    (c) Schematic of a three-channel silicon rib-waveguide array coated with \ch{Sb2Se3}. A focused laser locally writes a \(500\,\text{nm}\)-pixel pattern along a \(50\,\mu\text{m}\) section, enabling a dynamically reconfigurable optical coupling matrix.  
    (d) varFDTD simulations show intensity maps for three input ports that confirm an anti-diagonal coupling matrix with uniform phases.  
    (e) End-to-end inverse-design workflow: latent vectors \(\mathbf{z}_i\) are transformed into pixel patterns via a WGAN generator. A NN-based transmission predictor computes differentiable performance estimates, and gradients \(\partial\text{fitness}/\partial\mathbf{z}_i\) guide iterative updates of \(\mathbf{z}_i\) until convergence, minimizing a combined mean-squared error (MSE)  and phase loss.
    \textbf{(f,\,g) HiLAB: TO combined with VAE and BO.}  
    (f) A Vision-Transformer-based VAE encodes \(256\times128\) binary metasurface patterns into an eight-dimensional latent space, forming a smooth, fabrication-compatible manifold for optimization.
    (g) BO jointly explores the eight-dimensional latent geometry and the physical hyperparameters \(\{t_{1}, t_{2}, \Lambda_{y}\}\). Each proposed candidate is decoded, binarized, and evaluated with full-wave FDTD simulations. The optimization aims to maximize the worst-case diffraction efficiency across three wavelengths (470 nm, 550 nm, and 660 nm) using FoM: \(\mathrm{FoM}=\min\{\eta_{470}, \eta_{550}, \eta_{660}\}\). The Optimization progress is visualized through a 2D PCA (principal component analysis)  projection of sampled designs.
    \textbf{Permissions:}
    Panels~(a, b) \textbf{Reproduced under the terms of the CC BY 4.0 license.}\textsuperscript{\cite{kudyshev2020machine}} \textbf{Copyright 2020, The Author(s).}
    Panels~(c–e) \textbf{Reproduced under the terms of the CC BY 4.0 license.}\textsuperscript{\cite{radford2025inverse}} \textbf{Copyright 2025, The Author(s).}
    Panels~(f, g) \textbf{Reproduced under the terms of the CC BY 4.0 license.}\textsuperscript{\cite{marzban2025hilab}} \textbf{Copyright 2025, The Author(s).}}
    \label{fig:Generative}
\end{figure}
\noindent
Central to this approach are techniques such as variational autoencoders (VAEs)~\cite{wang2019clonal}, other dimensionality reduction and manifold learning methods~\cite{zandehshahvar2022manifold}, and generative adversarial networks (GANs)~\cite{Jiang2019GLOnet,Ma2019Probabilistic,Liu2018GANMetasurface}.
These models are typically trained on datasets comprising existing device geometries, paired with their corresponding optical responses.

\hl{Through training, the model learns a compact latent space in which each point corresponds to a unique, ideally fabrication-consistent geometry, as shown in the input-side branch of Figure~\ref{fig:evolution}. A generative model then maps a low-dimensional latent vector $\mathbf{z}$ to the full device layout. Far from being restrictive, this learned manifold serves as a powerful prior that regularizes the design search, particularly valuable in multi-objective problems where complete training data are unavailable~\cite{yin2023solving}. A latent space trained on a subset of objectives can still guide optimization toward global solutions while avoiding out-of-distribution regions where surrogate predictions become unreliable. In this way, the input-side representation constrains the search to physically meaningful regions, complementing output-side learning in hybrid frameworks.}

 Generative models~\cite{ma2021deep} are particularly prominent for design proposal and exploration within learned latent spaces. Note that in previous sections, although we put the emphasis on the WaveY-Net net model, GLOnet in Figure~\ref{fig:Inloop}(e, left) is an input-side generative model~\cite{Jiang2019GLOnet}. Figures~\ref{fig:Generative}(a,b) depict a conditional adversarial autoencoder (c-AAE) pipeline for designing nanopatterned antennas \cite{kudyshev2020machine}. In this system, the antenna topology, along with key geometric parameters like unit-cell size and spacer thickness, is encoded by an encoder network (E) into a compact 17-dimensional latent space (Figure~\ref{fig:Generative}(a)). A generator network (G) subsequently learns to produce realistic device patterns from vectors sampled from this latent space. A discriminator (D) keeps samples plausible and aligned with a chosen prior, yielding a physics-informed manifold. After training, the generator G is coupled with a pre-trained optical surrogate model (in this case, a visual geometry group (VGG)-based CNN predicting efficiency, as shown in Figure~\ref{fig:Generative}(b)) to enable rapid offline evaluation of proposed designs. The c-AAE's primary function here is to learn an efficient representation of the input device designs—the latent space. New designs are generated by sampling or manipulating points within this learned input manifold.

Another example, shown in Figures~\ref{fig:Generative}(c-e), employs a CNN-assisted  Wasserstein GAN (WGAN)~\cite{arjovsky2017wasserstein} for the inverse design of reconfigurable photonic waveguide arrays \cite{radford2025inverse}. The device is a three-channel silicon rib-waveguide coated with the phase-change material antimony selenide (Sb$_2$Se$_3$); laser writing changes the material state and the coupling between channels (Figure~\ref{fig:Generative}c). A WGAN generator is trained to produce candidate phase-change patterns (representing the device geometry, Figure~\ref{fig:Generative}(e), left panel) from latent vectors $z_i$. Incidentally, a NN-based simulator then rapidly predicts the resulting transmission matrix of the device (an example field distribution is shown in Figure~\ref{fig:Generative}(d)). The differentiability of this simulator allows for the computation of performance gradients with respect to the latent variables $z_i$, guiding their iterative refinement to meet a specific design target (Figure~\ref{fig:Generative}(e), right panel). In this workflow, the WGAN learns to generate valid and potentially high-performing input patterns (the phase-change material configurations). Optimization is performed efficiently within the WGAN's latent space, which is a learned, compressed representation of the design space.

Hybrid approaches that combine learned input representations with standard optimizers are becoming common. The HiLAB framework (Figures~\ref{fig:Generative}(f,g)) integrates a VAE with Bayesian optimization (BO) for metagrating design \cite{marzban2025hilab}. A Vision Transformer-based VAE is first trained to encode 256×128 binary metagrating patterns into a low 8-dimensional latent space (Figure~\ref{fig:Generative}(f)). This process establishes a smooth, potentially fabrication-constrained, manifold for subsequent optimization. A Bayesian optimizer then operates in this learned geometric latent space concurrently with optimizing a small set of continuous physical hyperparameters, such as layer thicknesses ($t_1$,$t_2$) and the lattice constant ($\Lambda_y$) (Figure~\ref{fig:Generative}(g)). Each point suggested by the Bayesian optimizer in this joint space is decoded by the VAE's decoder into a candidate metagrating design, which is then evaluated using full-wave EM simulations. Including training costs, this method enabled at least a ten-fold reduction in the number of simulations needed compared to a conventional TO, while achieving better performance to advance state-of-the-art. The subsequent optimization leverages this compressed latent space to efficiently search for designs that maximize a worst-case diffraction efficiency metric across multiple target wavelengths.

The creation of a latent space through input-side learning serves as a powerful learned prior over the design space. This prior implicitly encodes information about good or physically plausible geometries. For instance, the c-AAE in Figure~\ref{fig:Generative}(a) \cite{kudyshev2020machine} compresses designs into a 17-dimensional compressed design space, while the VAE in the HiLAB framework (Figure~\ref{fig:Generative}(f)\cite{marzban2025hilab}) establishes an 8-dimensional latent space described as a smooth, fabrication-constrained domain. Reducing the dimensionality may improve local optima. Optimization then proceeds by manipulating variables within these structured latent spaces, as seen in Figure~\ref{fig:Generative}(e) \cite{radford2025inverse} and Figure~\ref{fig:Generative}(g) \cite{marzban2025hilab}. This prior guides the search, enhancing efficiency and increasing the likelihood of identifying valid, high-performing designs compared to unconstrained exploration in the full, high-dimensional parameter space. The structural properties of the latent space, such as smoothness or disentanglement of features, thus become critical factors for successful design outcomes.

Many input-side methods exhibit a beneficial modularity by decoupling the task of learning the design representation (e.g., via a VAE or GAN) from the task of evaluating the performance of a generated design. Performance evaluation can be handled by a separate surrogate model or a full physical solver. In the c-AAE pipeline (Figures~\ref{fig:Generative}(a,b)) \cite{kudyshev2020machine}, the auto-encoder proposes designs and a pre-trained surrogate predicts efficiency. Similarly, in the HiLAB framework (Figures~\ref{fig:Generative}(f,g) \cite{marzban2025hilab}), the VAE encodes and decodes designs, but full-wave FDTD simulations are used for their evaluation. Even in the WGAN-based approach for reconfigurable waveguides (Figures~\ref{fig:Generative}(c-e) \cite{radford2025inverse}) and the GLOnet approach (Figure~\ref{fig:Inloop} (e, left)), the NN predictors are distinct modules from the generators. This decoupling offers considerable flexibility, allowing for independent improvement or substitution of the generative model (the input representation learner) or the performance evaluator without necessitating a complete retraining of the entire system. For example, a more accurate or faster solver could be integrated at a later stage with an already trained generative model.

Input-side learning limits search to a learned manifold that mostly contains high-performing designs. Searching that manifold is easier than scanning the full high-dimensional space. It also aids transfer learning and reuse. A latent representation of device geometries—termed a \emph{shape manifold}~\cite{kiarashinejad2020deep,zandehshahvar2022manifold}, captures fundamental geometric features likely to remain relevant even when target physical parameters (e.g., the operational wavelength or polarization) change. As a result, far less data are needed to adapt a pre-trained model than to train one from scratch~\cite{An2019Objective}. The underlying scale-invariance of Maxwell's equations can contribute to this phenomenon~\cite{Jackson1998}, as characteristic feature sizes or patterns learned by the representation in one spectral regime often remain useful at other scales. Learning shared geometric features yields a more transferable knowledge base than output-side models, which are tied to specific conditions.
Once trained, generative models such as GANs and VAEs can sample or interpolate in the latent space to propose many novel designs. The c-AAE generator in Figures~\ref{fig:Generative}(a,b) \cite{kudyshev2020machine} learns to produce realistic device patterns, and the WGAN generator in Figure~\ref{fig:Generative}(e) \cite{radford2025inverse} produces candidate phase-change patterns. This capability helps escape local minima and uncover unconventional solutions.

\hl{Furthermore, input-side learning can be an enabler of knowledge discovery. While the latent variables in a trained generative model are not always directly interpretable, techniques from the broader ML community offer pathways to enhance their physical meaning. Disentanglement methods, for instance, aim to learn representations where each latent dimension corresponds to a single factor of variation in the data~\cite{higgins2017beta,chen2018isolating}. Frameworks like the $\beta$-TCVAE (Total Correlation VAE) explicitly penalize correlation between latent variables to encourage such factorial representations~\cite{chen2018isolating}. Applying these techniques to nanophotonic design could yield latent spaces where specific variables control distinct physical features (e.g., resonance location, feature size). This transforms the generative model from a \emph{black box} into an intuitive design tool, as has been demonstrated in related fields like materials discovery~\cite{iten2020discovering}, and substantiates its potential for knowledge discovery.}

Despite these strengths, input-side methods face hurdles. The first hurdle is the acquisition and creation of a sufficiently large, diverse, and high-quality training set of device designs and their associated performance metrics. Naive random sampling of the vast design space often results in a training dataset dominated by poor-performing structures. This leads to inefficient use of computational resources. This chicken-and-egg problem—whereby learning a good representation of high-performing designs ideally requires a dataset of such designs, which are often the very objects of the search—is a fundamental issue. To address this data scarcity, recent work uses hybrid data-generation schemes for training data generation. For instance, initial datasets might be seeded with designs found via early-stopped TO~\cite{marzban2025hilab} or through exploration with low-fidelity surrogate models~\cite{kim2024multi,mal2025maps}, thereby focusing the representation learning process on more promising regions of the design space. This highlights an inherent interconnectedness and potential for synergy between different classes of inverse design methods. Further challenges include the representational capacity and potential biases introduced by the chosen model architecture (VAE, GAN, etc.).
\section{Global Search Methods Accelerated by Representation Learning}
\label{sec:classeC}

\hl{Practical inverse-design problems are often under-constrained or evolve dynamically (e.g., intelligent metasurfaces with shifting objectives or fabrication limits)~\cite{lin2023enabling}. Representation learning mitigates these challenges through a powerful, two-pronged approach: (i) input-side models provide a reusable, physics-agnostic manifold of device geometries, enabling efficient exploration and generation of new design candidates when constraints change  (Section~\ref{sec:classeB}), while (ii) output-side models provide a fast surrogate solver, enabling rapid evaluation and re-optimization of those candidates against the new targets (Section~\ref{sec:classeA}). This transferability not only facilitates efficient adaptation to evolving requirements but also supplies the foundation for making subsequent global searches computationally tractable under strict simulation budgets. Leveraging these transferable representations, global optimization schemes such as} GAs~\cite{wu2024localized,ren2021wavelength}, BO~\cite{li2022bayesian,wray2024optical,garcia2021bayesian}, and RL~\cite{sajedian2019optimisation,park2024sample,li2023deep} aim to escape local optima by sampling or evolving designs across the feasible space. Unlike gradient-based design methods, which often converge to a local solution dependent on initial conditions, global methods systematically explore regions that gradient-based methods often miss. However, the main practical constraint is the computational budget required by these global optimization methods. Although theoretically capable of reaching global optima given infinite evaluation samples, methods such as GAs and RL quickly become computationally infeasible due to the necessity of full-wave EM simulations at each design evaluation. This computational cost scales exponentially~\cite{li2022empowering} with the dimensionality of the design space, limiting the number of iterations and thus the exploration depth within realistic research scenarios.

To address the computational challenges of global optimization, recent approaches combine these methods with representation learning, either by using Class A surrogates or Class B latent spaces within the global search. In this way, global optimization can directly leverage the advantages of reduced evaluation cost or lower-dimensional search space provided by these tools. Representation learning lowers cost in two directions. Class A surrogates replace the full solver, making each evaluation inexpensive. Class B methods learn a low-dimensional latent space, so the optimizer explores fewer variables. Retraining with selected simulation data improves their accuracy in promising regions of the design space~\cite{pestourie2020active}. These reduced-dimensional latent spaces enhance the efficiency of global exploration. HiLAB is one Class B example~\cite{marzban2025hilab}. As shown in Figure~\ref{fig:Generative}(g), a VAE is used to encode \(256 \times 128\) freeform metasurface patterns into an 8-dimensional latent space—achieving over 4,000-fold compression. This allows global optimization to effectively control complex geometries using only eight latent variables, which are jointly optimized with physical hyperparameters via BO. Because BO uses uncertainty-guided sampling, it needs few optimization steps, and hence only a limited number of full-wave EM simulations. This combination lowers computational cost through low-dimensional exploration.
\begin{figure}[H]
  \centering
  \includegraphics[width=\linewidth]{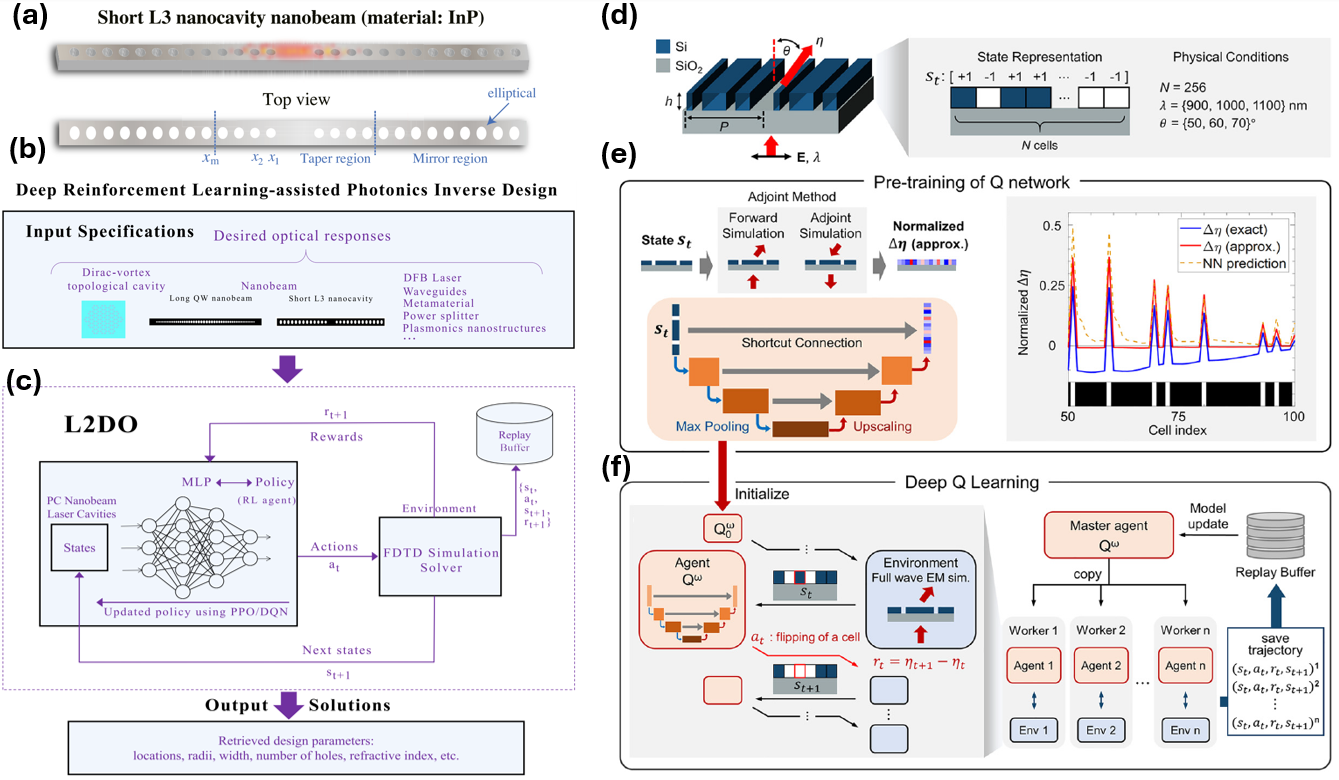}
\caption{\textbf{RL workflows for hybrid inverse design.}\\
  \textbf{(a–c) L2DO nanocavity synthesis.}
  (a) Short L\textsubscript{3} InP nanobeam cavity with symmetric taper and mirror holes
  \((x_{1},x_{2},\ldots ,x_{m})\).
  (b) User-specified optical targets are fed to the \emph{L2DO} engine.
  (c) Deep-RL loop: a four-layer MLP policy (PPO/DQN) interacts with an FDTD environment; replay-buffered
  experience tuples \((s_{t},a_{t},r_{t},s_{t+1})\) guide optimization of hole positions, radii, and counts.
  \textbf{(d–f) Physics-informed RL (PIRL) for metagrating optimization.}
  (d) Binary state encoding of a Si/SiO\(_2\)(Si: silicon, SiO\(_2\): silicon oxide) meta-grating; objective—maximize first-order TM deflection efficiency
  \(\eta\).
  (e) Physics-informed pre-training: a U-Net learns sensitivity maps
  \(\Delta\eta_{\mathrm{approx}}\) from adjoint analysis (plot compares exact, adjoint-approximate, and NN-predicted
  \(\Delta\eta\)).
  (f) Parallel Deep-Q stage: the pretrained agent \(Q_{0}^{\omega}\) is cloned into 16 workers running full-wave
  simulations; trajectories populate a global replay buffer while the master network \(Q^{\omega}\) is synchronously
  updated, enabling efficient, large-scale exploration of the metagrating design space.
    \textbf{Permissions:}
    Panels (a–c) \textbf{Reproduced under the terms of the CC BY 4.0 license.}\textsuperscript{\cite{li2023deep}} \textbf{Copyright 2023, The Author(s).}
    Panels (d–f) \textbf{Reproduced under the terms of the CC BY 4.0 license.}\textsuperscript{\cite{park2024sample}} \textbf{Copyright 2024, The Author(s).}}
    \label{fig:RL_pipelines}
\end{figure}
Another Class B approach is the L2DO (Learning to Design Optical Resonators) framework~\cite{li2023deep}, illustrated in Figures~\ref{fig:RL_pipelines}(a-c). L2DO operates directly on the design space and learns a policy for navigating it. Although it does not construct an explicit latent manifold, the RL agent implicitly captures a prior over high-performing input configurations through interaction. In Figure~\ref{fig:RL_pipelines}(a), the target is a short linear three-hole defect cavity (L3)~\cite{kurdi2008quality} with elliptical air holes in an indium phosphide (InP) substrate. Figure~\ref{fig:RL_pipelines}(b) shows how optical targets guide the agent's exploration via full-wave FDTD simulations. As illustrated in Figure~\ref{fig:RL_pipelines}(c), a multilayer perceptron (MLP) policy network selects discrete actions, such as shifting or resizing air holes, based on the current design state. These actions are simulated, and the resulting FoMs (e.g., quality factor, wavelength, modal volume) are used as rewards to improve the policy. Over time, the agent learns an implicit input-side representation without requiring gradient access or surrogate models. L2DO is therefore an RL-driven Class B method in which the design space itself is the learning domain.

Figures~\ref{fig:RL_pipelines}(d-f) show the physics-informed RL (PIRL) framework adapted from~\cite{park2024sample}, which integrates physics-based surrogate modeling into RL for optimizing freeform photonic devices. The design task involves a one-dimensional metagrating composed of 256 binary cells (Si or air), with the goal of maximizing first-order transverse magnetic (TM) deflection efficiency. As shown in Figure~\ref{fig:RL_pipelines}(e), a U-Net is pre-trained to predict the efficiency change that would result from flipping the material in each design cell. These targets are computed using adjoint sensitivity analysis, which estimates how a small perturbation in the refractive index at each cell affects the overall device efficiency. Specifically, the efficiency change is approximated as $\Delta\eta_{\mathrm{approx}} = (\partial \eta / \partial n_i)\Delta n$, where $\Delta n \approx 2.5$ is the refractive index contrast between Si and air. This allows the U-Net to learn a surrogate model of the physical response (Class~A representation), mapping full device layouts to cell-level sensitivity profiles. Notably, the U-Net does not represent geometry in a compressed latent space; rather, it predicts how performance responds to local changes across the structure. The trained surrogate then initializes the Q-function (the expected reward for each action-state pair~\cite{sutton1998reinforcement}) in the RL agent (Figure~\ref{fig:RL_pipelines}(f)), which applies deep Q-learning to explore the design space. Starting from this physics-informed prior improves sample efficiency over uninformed RL. This example shows how gradient-based surrogates can scale global optimization in high-dimensional inverse design settings, especially when simulation budgets are limited. Future work can integrate global optimizers with representation learning. For instance, coupling fast surrogate models (Class A) or robust latent-space representations (Class B) directly with powerful global methods like BO or RL could offer accelerated convergence toward practically achievable global optima. These hybrids methods navigate complex spaces yet stay within realistic compute budgets.

An alternative to large-scale global search is to use duality theory to derive provable limits on scattering, absorption, or near-field responses~\cite{chao2025bounds,molesky2025inferring,strekha2023suppressing}.  
By relaxing the inverse design problem into a convex dual program, one obtains upper bounds that certify how close any candidate geometry is to the physical optimum.  
In practice, the dual solution often suggests near-optimal field profiles or material distributions, which can be converted into a good initial guess for local optimization or TO.  
Thus, duality can (i) avoid wasting simulations on regions that cannot beat the bound and (ii) provide well-informed starting points that accelerate convergence, circumventing the heavy simulation cost faced by GA or RL methods.  
Recent demonstrations include bound-guided discoveries of strong light confinement~\cite{chao2023maximum} and multi-resonant devices~\cite{chao2022physical}, highlighting duality’s potential to replace the computationally expensive global search.

\section{Concluding Remarks}
\label{sec:future}

The field of nanophotonic inverse design has moved from simple parameter sweeps to hybrid pipelines that integrate physics-based algorithms, ML-driven exploration, and emerging numerical methods. While these developments have already facilitated high-performance devices, several directions deserve attention.

One concern is \emph{how to handle} and calibrate design complexity. Freeform topologies offer a large configuration space that can, in principle, yield superior performance at the risk of high computational costs and increased susceptibility to manufacturing errors~\cite{kang2024large}. More constrained parameterizations, such as shape-based~\cite{dainese2024shape} or level-set methods~\cite{vercruysse2019analytical}, reduce the search space. However, they potentially sacrifice some bandwidth or multi-resonant behavior. Future research may develop rigorous metrics or \emph{complexity bounds} that quantify when adding more degrees of freedom ceases to provide meaningful gains, enabling more systematic choices about the granularity of optimization with the simplest parameterization that does not sacrifice performance.

\emph{Data availability}\hl{ remains a critical bottleneck. To mitigate the high cost of full-wave simulations, hybrid data generation strategies are essential. Multi-fidelity training, which leverages numerous cheap, low-fidelity simulations (e.g., from approximate solvers) to inform a model trained on a few high-fidelity examples, is a promising direction~\cite{liu2018zongfu,lu2022multifidelity}. Similarly, active learning can intelligently select the most informative designs to simulate next, drastically reducing the data required to train an effective surrogate model~\cite{Ma2019Probabilistic,pestourie2020active}. These strategies, coupled with community efforts to standardize and share datasets like MetaNet~\cite{jiang2020metanet}, are crucial for progress.}

\hl{This focus on data generation leads to a deeper methodological question regarding its use: the utility of a physics-informed loss when training on data already generated by a high-fidelity solver. While such a loss is the central element for frameworks like MaxwellNet~\cite{lim2022maxwellnet} that learn directly from the PDE, its added value in purely data-driven contexts is more debatable. The effectiveness of a physics-informed loss depends on the dimensionality and composition of the design space. As highlighted in~\cite{pestourie2025fast}, two main types of degrees of freedom govern the problem: physical variables governed by Maxwell’s equations and geometrical parameters derived from a choice of parameterization. When the number of physical degrees of freedom is small and the training data already comply with the governing equations, adding an explicit physics term offers limited benefit. In contrast, when both physical and geometrical degrees of freedom are large and strongly coupled, where the curse of dimensionality becomes significant, the inclusion of a PDE-based constraint provides a useful inductive bias that promotes physically consistent learning and improved generalization~\cite{chen2022high}.}

Another goal is to build representations that work across geometry choices and length scales. Most current models are trained for a single size or operating band and must be retrained for a new scale. The goal is to learn underlying design principles that are inherently scalable, enabling knowledge transfer from optimizing, for example, a $100\,\text{\textmu m}$ diameter device directly to a $500\,\text{\textmu m}$ counterpart. This would enhance the transferability and reusability of learned representations, reducing reliance on entirely new datasets and training campaigns for each variation in device dimensions. With size-aware features, a pre-trained model could be adapted with minimal fine-tuning, saving data and computational resources.
\hl{To this end, several emerging research directions, extending beyond the latent-space approaches of Section~\ref{sec:classeB}, seek to develop representations that are both transferable and scalable across geometry choices and length scales. A major frontier is explicit transfer learning, where knowledge gained from a source task is adapted to a new target system. This paradigm is being explored through diverse strategies, including cross-domain and heterogeneous learning frameworks that aim to bridge distinct physical domains or structural families~\cite{zhang2022heterogeneous, gao2024cross, peng2024transfer}. Complementing these efforts, mesh-free formulations are drawing increasing attention to overcome discretization dependence. Notably, geometric neural parameterization employs neural networks to learn continuous, mesh-independent mappings of device boundaries, enabling smoother generalization across geometric variations~\cite{dai2025shaping}. At a broader level, operator-learning frameworks such as the Fourier Neural Operator~\cite{wu2025data,augenstein2023neural,gu2022neurolight} pursue generalization by directly learning the underlying PDE solution operator, thereby constructing surrogates that can adapt to new physical conditions without retraining from scratch. Graph Neural Networks can encode relational and topological features that remain invariant under geometric scaling, a property particularly valuable for learning design principles transferable across device sizes and configurations~\cite{khoram2022graph,kuhn2023exploiting,yan2022all}. Together, these approaches point toward a new generation of models that aim to capture reusable, physics-aware design representations, though determining which of them will ultimately prove most effective remains an open challenge for the field.

Many of these advanced strategies succeed by embedding desirable mathematical properties directly into the learned representations to yield better-behaved optimization landscapes.} Autoencoders, for example, cut the effective dimension, which often improves local minima. Furthermore, neural representations provide continuous relaxations for inherently discrete or binary design problems, such as TO, so gradient descent can still be leveraged even when the final design must be binary. 

Disentangling latent variables, so each maps to a clear physical feature, can smooth the landscape, let designers move along or sample meaningful axes, and help escape local minima. Insights from optimal transport~\cite{liu2018latent} or manifold learning~\cite{zandehshahvar2022manifold} could inform representations that are both compact and conducive to robust optimization.

\hl{A practical challenge is ensuring optimized structures are manufacturable. Future work should move beyond post-design checks and bake fabrication constraints into the learned representation itself. For instance, generative models could be inspired by methods that combine electromagnetic and structural topology optimization to produce self-supported, 3D-printable designs~\cite{hammond2021photonic}, or by level-set methods that inherently produce smooth, fabricable boundaries~\cite{hammond2025unifying, hiesener2025seeded}.}

\hl{Finally, advancing beyond single-physics optimization to tackle coupled multiphysics represents a major research frontier. Future devices will require co-design across domains like electromagnetics, thermodynamics, and mechanics. Promising frameworks are emerging to tackle this challenge. For instance, advanced formulations of PINNs are being developed specifically to solve tightly coupled thermo-mechanical systems~\cite{harandi2024mixed}. In parallel, material-informatics-based inverse design has been used to efficiently co-optimize metamaterials for distinct physical objectives, such as achieving selective performance in both the visible and infrared spectra~\cite{xi2023ultrahigh}. While still an emerging area, these pioneering works demonstrate the high potential of multiphysics inverse design for creating truly multifunctional, robust devices.}

\hl{Further challenges include the representational capacity of the chosen model architecture. This is a particularly critical consideration with the rise of \emph{multifunctional} devices, which inherently involve \emph{multi-objective} optimization where the goal is to map out a \textbf{Pareto} \textbf{front} of optimal trade-offs~\cite{deb2011multi,deb2002fast}. A single design might excel in one objective (e.g., high efficiency at one wavelength) at the expense of another, while a different design might offer balanced performance across all objectives; both are valid solutions along the front. Since generating a training set that uniformly covers this entire front is often computationally prohibitive, the data is frequently biased toward specific, high-interest regions. Therefore, the \emph{coverage} of the learned manifold becomes a strategic element of the design process rather than a simple shortcoming~\cite{jiang2020multiobjective,whiting2020meta,manfredi2024stochastic}. This reality, however, raises critical questions for future work. For example, can \emph{transfer} \emph{learning} be used to leverage a model trained on one region of the Pareto front to efficiently explore another? Furthermore, it remains an open question whether a \emph{single} or \emph{shared} \emph{latent} \emph{representation} is sufficient to capture the complex topology of an entire front, or if more sophisticated, task-specific representations are necessary. Finally, a related challenge is that of \emph{interpretability}; while some latent dimensions may correlate with intuitive physical features, achieving a comprehensive understanding of all learned variables and their relationships is not guaranteed. Addressing these questions is a key frontier for the field.}

Progress on the numerical front is also reshaping inverse design. Faster integral-equation solvers \cite{hu2021chebyshev,garza2022fast}, GPU-accelerated PDE codes \cite{paul2025accelerated,um2020solver,bhatia2025prdp}, and symmetry-aware solvers \cite{sun2025scalable,christiansen2020fullwave} are cutting simulation time. These faster solvers can mesh with iterative or hybrid algorithms and enlarge the tractable region of the search space, even if an exhaustive brute-force sweep remains out of reach. Frameworks that embed physical laws, such as Maxwell’s equations, directly into ML architectures as a solver layer may become practical to keep model outputs physically valid.

\hl{A central challenge in inverse design is the one-to-many problem, where multiple distinct designs yield the same output. While generative input-side models like cGANs can address this by producing a diverse family of plausible solutions, they are inherently limited to the scope of their training data. Hybrid frameworks provide powerful strategies to transcend this limitation and discover higher-performance or novel solutions through two key synergies.
First, pairing an input-side model with a differentiable output-side surrogate enables gradient-based local refinement~\cite{radford2025inverse,chen2022high,jing2022neural}. A design generated from the latent space serves as a high-quality starting point, which is then fine-tuned using gradients from the surrogate to converge to a superior local optimum. Second, coupling an input-side latent space with a global search algorithm is uniquely powerful for co-optimizing the device pattern alongside its physical parameters (e.g., layer thicknesses, periodicity)~\cite{marzban2025hilab,Jiang2019GLOnet,tian2024high}. While a generative model can provide diverse patterns for a fixed engineering objective, a global optimizer on a low-dimensional latent space can efficiently explore the much larger joint space of geometry and physical dimensions. This is critical in photonics for discovering novel resonances that emerge from the interplay between pattern and structure, a task for which standalone generative models are ill suited ~\cite{jiang2025near,elsawy2019global}. This transforms the design process from a static selection of plausible patterns to a dynamic search for truly optimal, and often non-intuitive, devices.}

In practice, selecting the best hybrid strategy for a given problem remains non-trivial. Common metrics for evaluating performance, such as objective function values or peak efficiencies, do not always capture robustness, fabrication constraints, or interpretability. Moreover, systematic guidelines for combining different modules (e.g., deciding when to switch from global search to local refinement) are still missing, highlighting the need for more research. \hl{At the heart of this strategic challenge is the trade-off between the input-side prior and the output-side physics constraint. A highly regularized input-side prior can relax the accuracy requirements of the output-side surrogate, while a highly accurate surrogate enables a broader, less constrained search. Navigating the balance between these complementary forms of regularization is a fundamental consideration for the future of advanced inverse design.}
 
\hl{Looking ahead, it is clear that hybrid inverse design approaches, uniting global exploration, local refinement, advanced solvers, and data-driven modeling, will continue to impact the development of nanophotonic devices. This integration allows designers to balance key trade-offs: geometric freedom versus manufacturing constraints, computational cost versus multi-functional objectives, and empirical performance versus fundamental physical limits. The ultimate expression of this synergy may be the emergence of autonomous multi-agentic frameworks that leverage these hybrid methods as a high-performance computational backend. Such frameworks can translate high-level, semantic design goals into optimized device layouts in a nearly real-time, automated fashion, fundamentally shifting the paradigm of scientific discovery~\cite{lupoiu2025multi}. Crucially, by enabling the optimization of specific near-field distributions and optical modes, these advanced frameworks move beyond simple performance maximization, providing deeper insights into the underlying physics of light-matter interactions~\cite{jiang2025near}. The end result will be a more profound understanding of how to co-engineer light and matter at the nanoscale.}

\section*{Supporting Information}
Correspondence and requests for materials should be addressed to the authors.
\medskip
\section*{Acknowledgements}
This work was supported by National Science Foundation.  
\bibliographystyle{unsrt}
\bibliography{Mybib}

\end{document}